\documentclass[conference]{IEEEtran}
\IEEEoverridecommandlockouts
\usepackage[square, comma, sort&compress,numbers]{natbib}
\usepackage{amsmath,amssymb,amsfonts}
\usepackage{algorithmic}
\usepackage{graphicx}
\usepackage{textcomp}
\usepackage{xcolor,soul}
\usepackage{url}
\usepackage{hyperref}
\usepackage{tabularx}
\usepackage{booktabs}
\usepackage{multicol}
\usepackage{multirow}
\usepackage[caption=false]{subfig}
\usepackage{float}

\def\BibTeX{{\rm B\kern-.05em{\sc i\kern-.025em b}\kern-.08em
    T\kern-.1667em\lower.7ex\hbox{E}\kern-.125emX}}
\newcommand\norm[1]{\left\lVert#1\right\rVert}
    
\begin{document}

\title{FloLPIPS: A Bespoke Video Quality Metric for Frame Interpolation\\
\thanks{This work was funded by the China Scholarship Council, University of Bristol, and the UKRI MyWorld Strength in Places Programme (SIPF00006/1).}
}

\author{\IEEEauthorblockN{Duolikun Danier, Fan Zhang and David Bull}
\IEEEauthorblockA{Visual Information Laboratory \\
University of Bristol\\
Bristol, BS1 5DD, United Kingdom\\
\{Duolikun.Danier, Fan.Zhang, Dave.Bull\}@bristol.ac.uk}
}

\maketitle


\begin{abstract}
Video frame interpolation (VFI) serves as a useful tool for many video processing applications. Recently, it has also been applied in the video compression domain for enhancing both conventional video codecs and learning-based compression architectures. While there has been an increased focus on the development of enhanced frame interpolation algorithms in recent years, the perceptual quality assessment of interpolated content remains an open field of research. In this paper, we present a bespoke full reference video quality metric for VFI, FloLPIPS, that builds on the popular perceptual image quality metric, LPIPS, which captures the perceptual degradation in extracted image feature space. In order to enhance the performance of LPIPS for evaluating interpolated content, we re-designed its spatial feature aggregation step by using the temporal distortion (through comparing optical flows) to weight the feature difference maps. Evaluated on the BVI-VFI database, which contains 180 test sequences with various frame interpolation artefacts, FloLPIPS shows superior correlation performance (with statistical significance) with subjective ground truth over 12 popular quality assessors. To facilitate further research in VFI quality assessment, our code is publicly available at \url{https://danier97.github.io/FloLPIPS}.
\end{abstract}

\begin{IEEEkeywords}
Video Quality Assessment, Video Frame Interpolation, FloLPIPS.
\end{IEEEkeywords}

\section{Introduction}

Video frame interpolation (VFI) has recently attracted significant interest in the video compression research community as a means of generating intermediate frames between every two consecutive frames in a sequence. It has been employed in video coding to replace motion prediction in conventional codecs~\cite{8744618}, to perform error concealment~\cite{7423765}, and as the basis of end-to-end deep video compression systems~\cite{wu2018video}.

Existing works on VFI largely focus on developing new algorithms to improve interpolation performance under various challenging scenarios. These techniques include the use of deformable convolution~\cite{lee2020adacof, danier2022st}, transformer networks~\cite{shi2022video, lu2022video}, coarse-to-fine architectures~\cite{chen2021pdwn, danier2022enhancing, kong2022ifrnet}, and the design of more flexible optical flow estimation mechanisms~\cite{niklaus2020softmax, hu2022many}. While a plethora of VFI algorithms have been reported, the perceptual evaluation of frame interpolated content has attracted less attention. The most common quality assessment methods used for VFI are PSNR, SSIM~\cite{wang2004image}, and LPIPS~\cite{zhang2018unreasonable}. However, these metrics have recently been shown exhibit poor correlation with subjective opinion scores for the case of frame interpolated videos \cite{danier2022subjective}. Other popular image/video quality assessment models developed for more generic applications, such as VMAF~\cite{li2016toward}, VIF~\cite{sheikh2005information}, as well as the ones that specifically consider frame rate-related artefacts including ST-GREED~\cite{madhusudana2021st} and FRQM~\cite{zhang2017frame}, have also failed to provide satisfactory performance in the context of VFI~\cite{danier2022subjective}. Due to the lack of a high performance quality metric, many VFI works~\cite{liu2017video, kalluri2020flavr, danier2022st} resort to performing costly subjective experiments to evaluate the visual quality of their interpolation results. Hence, there is an urgent need to develop a quality assessment model that can more accurately capture the perceptual degradation in frame interpolated videos.


In this context, we propose a full reference video quality assessment model, Flow difference-weighted LPIPS, FloLPIPS, which, to the best of our knowledge, is the first bespoke metric specifically designed for video frame interpolation. FloLPIPS combines the spatial distortion captured by LPIPS with the temporal degradation estimated using the discrepancy between the optical flow maps of the reference and distorted videos. 
Our experiments show that FloLPIPS achieves state-of-the-art performance on a publicly available subjective database for VFI, significantly outperforming 12 other tested quality metrics. 

The rest of the paper is organised as follows. We first describe the FloLPIPS algorithm in Section~\ref{sec:method}. The quantitative evaluation results and analysis of the proposed method are then presented in Section~\ref{sec:results}. Finally, we draw conclusions in Section~\ref{sec:conclusion}.

\begin{figure*}[t]
    \centering
    \includegraphics[width=\linewidth]{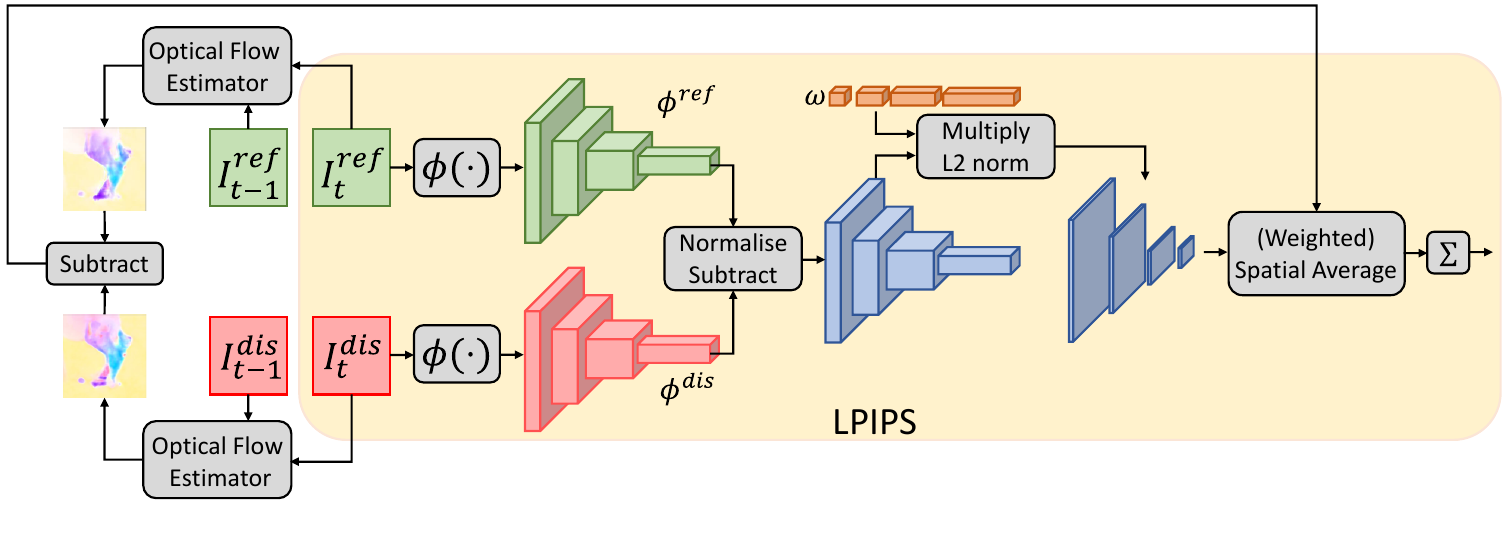}
    \vspace{-3.5em}
    \caption{The overall pipeline of the proposed FloLPIPS. The part in pale yellow corresponds to the original LPIPS. Here we additionally capture signal degradation in the temporal domain by computing the optical flows between frames. The difference between the reference and distorted optical flows are used to weight the feature maps before performing spatial averaging in LPIPS.}
    \label{fig:overall}
\end{figure*}

\section{Proposed Method}\label{sec:method}
The workflow to calculate the proposed Flow difference-weighted LPIPS, FloLPIPS, is illustrated in Fig.\ref{fig:overall}. This comprises two primary stages: (i) the calculation of LPIPS features and (ii) flow difference based feature aggregation.


The Learned Perceptual Image Patch Similarity (LPIPS) method measures the perceptual distortion between two images based on the feature maps extracted by several early layers of a CNN (e.g. five for VGG~\cite{Simonyan15} and AlexNet~\cite{krizhevsky2012imagenet}, and seven for SqueezeNet~\cite{iandola2016squeezenet}), which is pre-trained for image classification. Specifically, given a reference and a distorted frame, $I^{ref}, I^{dis}$, a feature extractor $\phi(\cdot)$ extracts feature maps $\phi_l^{ref}$ and $\phi_l^{dis}$ from these two frames at multiple layers, where $l=1,2,...,L$ and $L$ is the total number of layers employed. The reference and distorted feature maps are then normalised, and their differences are calculated. Next, the difference maps are weighted linearly at each layer by learned vectors $\omega_l$, and the $\ell2$ norm is obtained across the channel dimension. 

In the original LPIPS, the resulting feature difference of each layer is averaged spatially and summed to obtain the final LPIPS score, shown in (\ref{eqn:l2}).
\begin{equation}
    \mathrm{LPIPS}(I^{ref}, I^{dis}) = \sum_{l} \frac{1}{H_l W_l} \sum_{h,w} \norm{w_l\odot (\phi_{l,hw}^{ref} - \phi_{l,hw}^{dis})}^2_2\label{eqn:lpips}
\end{equation}
where $H_l,W_l$ are the height and width of the feature maps at layer $l$, and where $h,w$ index over all the spatial locations. The symbol $\odot$ represents a Hadamard product. More details on LPIPS training can be found in the original paper~\cite{zhang2018unreasonable}.

It is noted that, although LPIPS offers promising performance for measuring the perceptual quality of images, we previously reported that it exhibits unsatisfactory correlation with human judgement when used to evaluate the quality of interpolated videos~\cite{danier2022subjective}. We observed two major reasons for this. Firstly, LPIPS only captures distortions in the spatial domain, without any measure of temporal consistency. Secondly, when aggregating the spatial information, an arithmetic mean is taken over the feature difference maps where each pixel is weighted equally. This conflicts with VFI algorithms which tend to introduce salient artefacts in regions with movement which have a non-uniform spatial distribution. 

Based on these observations, instead of considering single reference and distorted frames, we additionally involve the previous frames (both for reference and distorted sequences) to better capture the temporal characteristics of the video. Specifically, we compute the optical flow $F^{ref}$ between the reference frames $I^{ref}_{t-1}$ and $I^{ref}_t$, as well as the flow $F^{dis}$ between the corresponding distorted frames $I^{dis}_{t-1}$ and $I^{dis}_t$. To measure the distortion in the temporal domain, we compute the magnitudes of the differences between the reference and distorted flow maps. This difference, after normalisation, is used to weight the feature difference maps obtained during the calculation of LPIPS, so that a weighted spatial average is computed instead of the original arithmetic mean. Such a spatial pooling strategy places more emphasis on those pixels where there is more motion discrepancy. The underlying assumption is that the regions with distorted motions correspond to more salient parts of the video. This process is summarised in (\ref{eqn:flo1})-(\ref{eqn:flolpips}) below.
\begin{equation}
    F^{ref} = \mathrm{OpticalFlow}(I^{ref}_{t-1}, I^{ref}_{t})\label{eqn:flo1}
\end{equation}
\begin{equation}
    F^{dis} = \mathrm{OpticalFlow}(I^{dis}_{t-1}, I^{dis}_{t})
\end{equation}
\begin{equation}
    \Delta F = \norm{F^{ref} - F^{dis}}_2 \label{eqn:l2}
\end{equation}
\begin{equation}
    \Delta \hat{F} = \frac{\Delta F}{\sum_{h,w}\Delta F_{hw}}
\end{equation}
\begin{gather}
    \mathrm{FloLPIPS}(I^{ref}_{t-1}, I^{ref}_{t},I^{dis}_{t-1}, I^{dis}_{t}) =\quad\quad\quad\quad\quad\quad\quad\quad\quad\quad\quad \nonumber \\
    \sum_{l} \frac{1}{H_l W_l} \sum_{h,w}\Delta \hat{F}_{hw} \norm{w_l\odot (\phi_{l,hw}^{ref} - \phi_{l,hw}^{dis})}^2_2\label{eqn:flolpips}
\end{gather}
The optical flow estimator used is a pre-trained PWC-Net~\cite{sun2018pwc}, and a pre-trained AlexNet (the first five layers) is used for LPIPS feature extraction. It is noted that all the model and hyper-parameters in the proposed FloLPIPS are identical to those in the original LPIPS or in the pre-trained PWC-Net and AlexNet. No other parameters need to be further optimised for this approach.

In order to obtain the quality score for the whole video, FloLPIPS is calculated for every two frames in a sliding window with a stride of 1, and all the frame level scores are averaged. 

\section{Results and Discussion}\label{sec:results}
In this section, we first describe the experimental setup, then quantitatively evaluate the performance of FloLPIPS by comparing it with 12 commonly used image/video quality metrics. Ablation study results are also presented to demonstrate the efficacy of the proposed flow difference-based weighting.

\subsection{Experimental Setup}
\textbf{Database}. We use the BVI-VFI~\cite{danier2022subjective} database to evaluate the proposed method, as it is the only publicly available subjective quality database that contains uncompressed video sequences with only VFI-induced distortions. BVI-VFI includes 36 reference sequences with a spatial resolution of 1920$\times$1080 at 30, 60 and 120fps. Five VFI methods were used to generate 180 distorted sequences, and these methods cover several major classes of VFI, including basic frame averaging/repeating, purely flow-based deep learning methods (DVF~\cite{liu2017video} and QVI~\cite{xu2019quadratic}), as well as more recent kernel-based state of the art (ST-MFNet~\cite{danier2022st}). For each distorted sequence, BVI-VFI provides a ground-truth Differential Mean Opinion Score (DMOS), which represents the perceptual quality difference between the distorted and reference sequences.

\textbf{Compared Methods}. The three commonly used metrics in current VFI research, namely PSNR, SSIM~\cite{wang2004image}, and LPIPS~\cite{zhang2018unreasonable} are included for comparison. In addition, we evaluate several other popular image quality metrics: MS-SSIM~\cite{wang2003multiscale}, VIF~\cite{sheikh2005information}, VSI~\cite{zhang2014vsi}, and also a more recent deep learning model CONTRIQUE~\cite{9796010}. Since these image quality models do not include temporal information, we also evaluate three generic video quality metrics, including ST-RRED~\cite{soundararajan2012video}, VMAF~\cite{li2016toward}, and C3DVQA~\cite{xu2020c3dvqa}, and two bespoke metrics that were designed to address frame-rate related distortions, ST-GREED~\cite{madhusudana2021st}, and FRQM~\cite{zhang2017frame}. For all learning-based methods, we have used the pre-trained versions released by the authors. For fair comparison, we only focus on full or reduced reference quality models in this experiment, since FloLPIPS is a full reference model.

\textbf{Evaluation Metrics}. We measure the quality metric performance based on three different statistical methods: Pearson's Linear Correlation Coefficient (PLCC), Spearman's Rank-Order Correlation Coefficient (SROCC) and Root Mean Squared Error (RMSE). For the computation of PLCC and RMSE, we first fit a logistic function between the calculated quality indices and the DMOS values according to \cite{video2000final}:
\begin{equation}
    Y(x) = \beta_2 + \frac{\beta_1 - \beta_2}{1 + \exp\big( -\frac{x-\beta_3}{|\beta_4|} \big)}
\end{equation}

\begin{table}[t]
\centering
\caption{The performance of the evaluated quality assessment models on the BVI-VFI dataset. For each statistical metric, the best and second best results are \textbf{bolded} and \underline{underlined} respectively. Runtime denotes the time taken to process a single 1920$\times$1080 frame.}
\label{tab:quantitative}
\begin{tabular}{c|c|c|c|c}
\toprule
Metric & PLCC $\uparrow$ & SROCC $\uparrow$ & RMSE $\downarrow$ & runtime (ms)\\
\midrule 
PSNR & 0.471 & 0.520 & 19.358 & 15.773 \\
SSIM & 0.475 & 0.581 & 19.328 & 199.107 \\
LPIPS & \underline{0.597} & 0.599 & \underline{17.603} & 59.153 \\
MS-SSIM & 0.529 & 0.593 & 18.623 & 282.500 \\
VIF & 0.489 & 0.535 & 19.152 & 6888.673 \\
VSI & 0.575	& \underline{0.631}	& 21.938 & 634.987 \\
CONTRIQUE & 0.545 & 0.309 & 18.400 & 256.047 \\
VMAF & 0.564 & 0.595 & 18.115 & 98.487 \\
C3DVQA & 0.351 & 0.508 & 20.936 & 153.507\\
ST-RRED & 0.568 & 0.610 & 18.063 & 1115.1\\
ST-GREED & 0.214 & 0.112 & 21.432 & 142.547 \\
FRQM & 0.456 & 0.535 & 19.525 & 67.953 \\
\midrule
FloLPIPS & \textbf{0.706} & \textbf{0.683}& \textbf{15.546} & 332.3 \\
\bottomrule	
\end{tabular}
\end{table}

\begin{table}[t]
\centering
\caption{F-test results between DMOS prediction residuals of selected quality metrics at a 95\% confidence interval. The value ``1'' indicates the metric in the row is superior to the metric in the column and ``-1'' means the opposite, while a ``0'' denotes statistical equivalence.}
\label{tab:ftest}
\resizebox{\linewidth}{!}{
\begin{tabular}{c|c|c|c|c|c|c}
\toprule
Metric & MS-SSIM & VMAF & LPIPS & ST-RRED & VSI & FloLPIPS \\
\midrule 
MS-SSIM   & - & 0 & -1 & -1 & 0 & -1 \\
VMAF   & 0 & - & -1 & -1 & 0 & -1 \\
LPIPS & 1 & 1 & - & 0 & 1 & -1 \\
ST-RRED  & 1 & 1 & 0 & - & 1 & -1 \\
VSI    & 0 & 0 & -1 & -1 & - & -1 \\
FloLPIPS   & 1 & 1 & 1 & 1 & 1 & - \\
\bottomrule	
\end{tabular}
}
\end{table}


\subsection{Quantitative Evaluation}

The evaluation of FloLPIPS alongside other compared quality models on the BVI-VFI dataset are presented in Table~\ref{tab:quantitative}. It can be observed that FloLPIPS outperforms the other 12 models according to all three performance measurements. Compared to its predecessor LPIPS, FloLPIPS achieves significant improvements in terms of PLCC and SROCC (+0.109 and +0.084 respectively). VSI and LPIPS provide the second best performance among all the tested methods based on different statistical metrics.

To validate the statistical significance of the superior performance of FloLPIPS, an F-test~\cite{seshadrinathan2010study} was performed between the prediction residuals (after non-linear fitting) of FloLPIPS and five other best-performing metrics (based on SROCC values); the results are shown in Table~\ref{tab:ftest}. Based on the F-test results, it can be confirmed that the performance improvement of FloLPIPS over the five best-performing quality metrics is statistically significant at the 95\% confidence interval. 

Table~\ref{tab:quantitative} also reports the runtime required by each quality metric to process a single 1920$\times$1080 frame. All algorithms were evaluated on an Intel Xeon W-1250 6-core CPU with 64GB RAM, and an NVIDIA RTX 3090 GPU was used for deep learning-based models. The results show that runtime of FloLPIPS is relatively moderate.

\begin{table}[t]
\centering
\caption{Ablation study results on different weighting methods.}
\label{tab:ablation}
\begin{tabular}{c|c|c|c}
\toprule
Weight type & PLCC $\uparrow$ & SROCC $\uparrow$ & RMSE $\downarrow$ \\
\midrule 
w/o weight & 0.597 & 0.599 & 17.603 \\
reference flow & 0.627 & 0.648 & 19.398\\
distorted flow & 0.698 & 0.664 & 16.698 \\
difference & \textbf{0.706} & \textbf{0.683} & \textbf{15.546} \\
\bottomrule	
\end{tabular}
\end{table}

\begin{figure}[t]
    \centering
    \subfloat[Reference]{\includegraphics[width=0.495\linewidth]{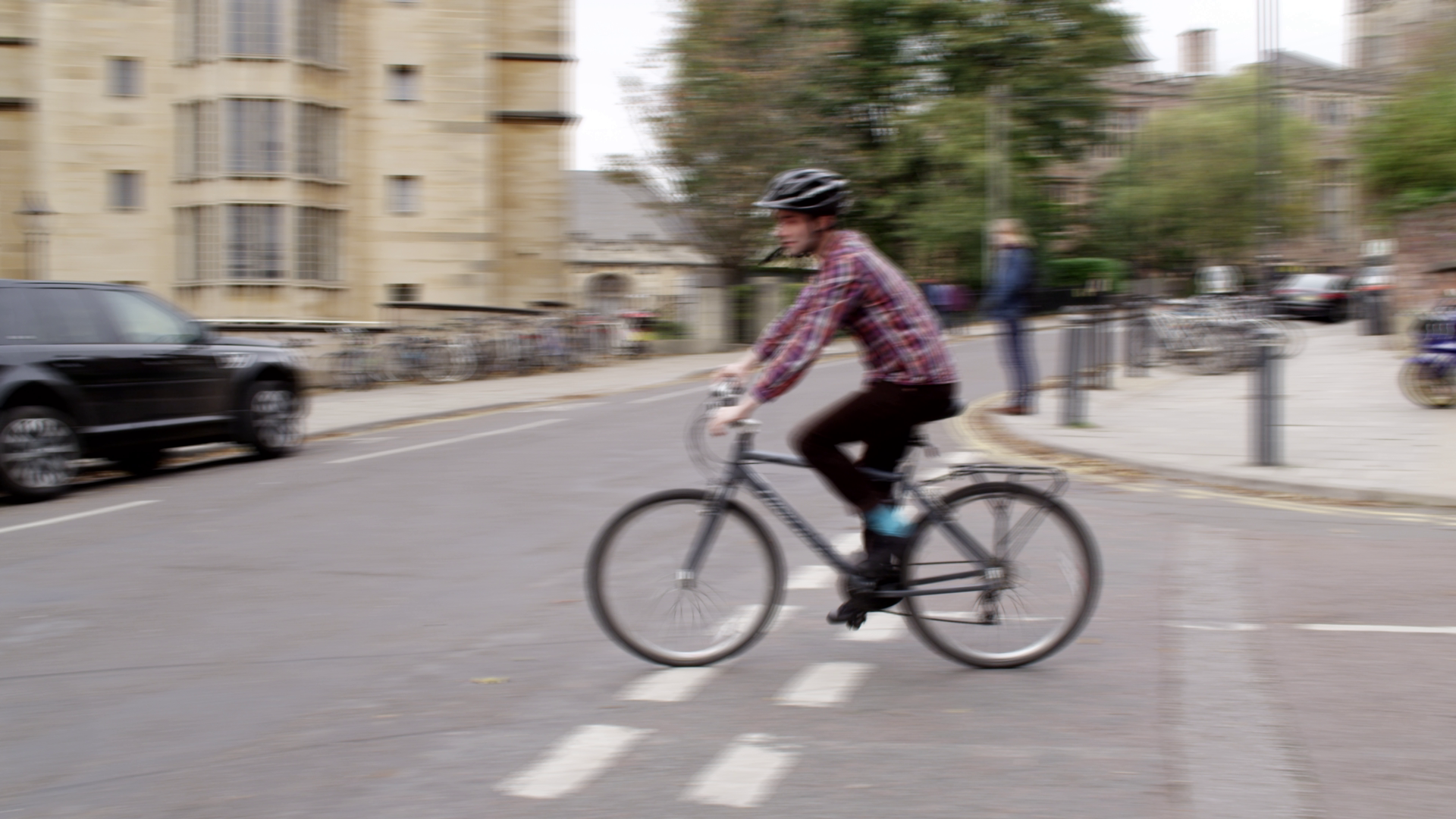}}\;\!\!
    \subfloat[Distorted]{\includegraphics[width=0.495\linewidth]{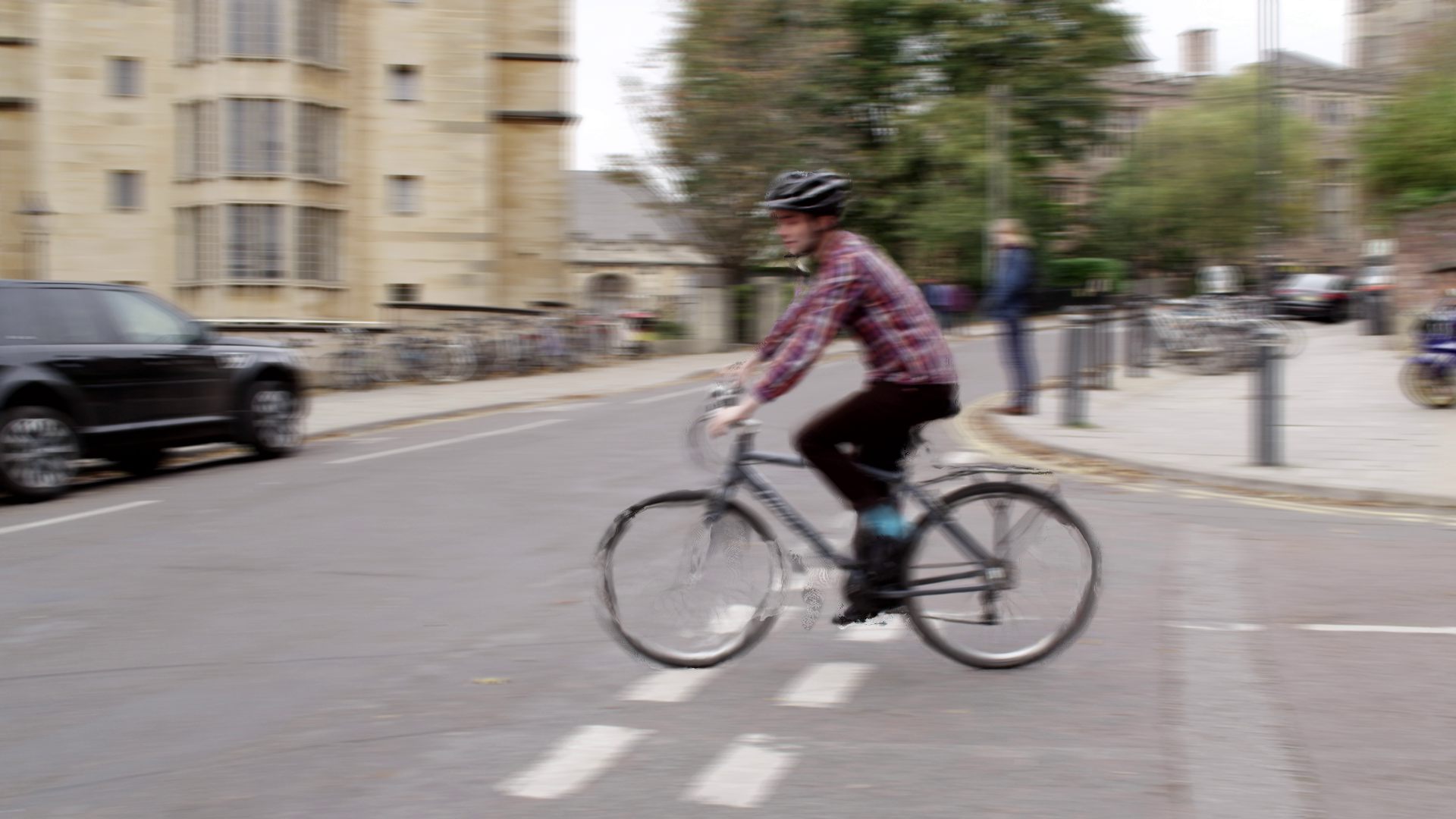}}\\
    \subfloat[Ref. flow weight]{\includegraphics[width=0.330\linewidth]{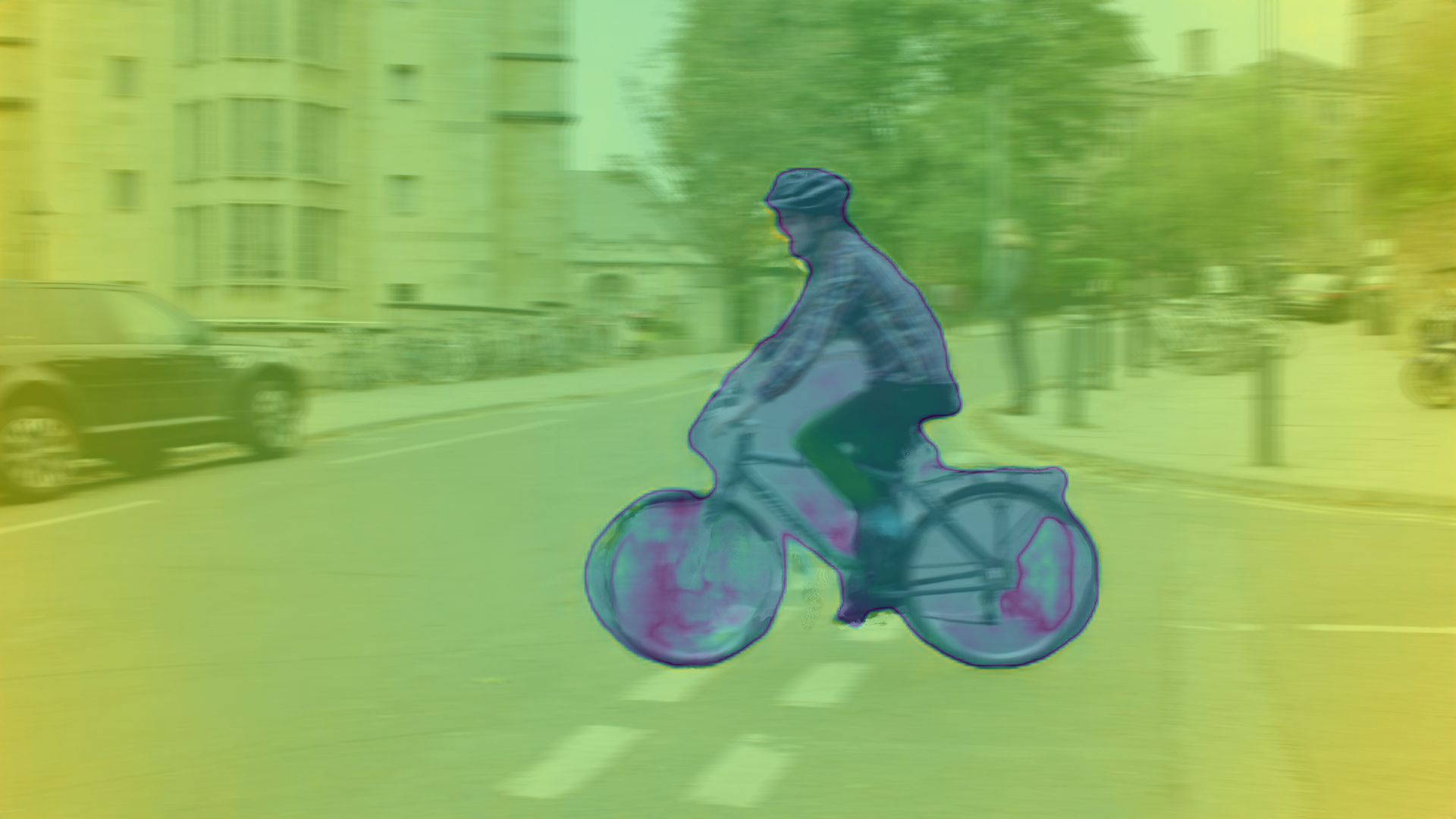}}\;\!\!\!
    \subfloat[Dis. flow weight]{\includegraphics[width=0.330\linewidth]{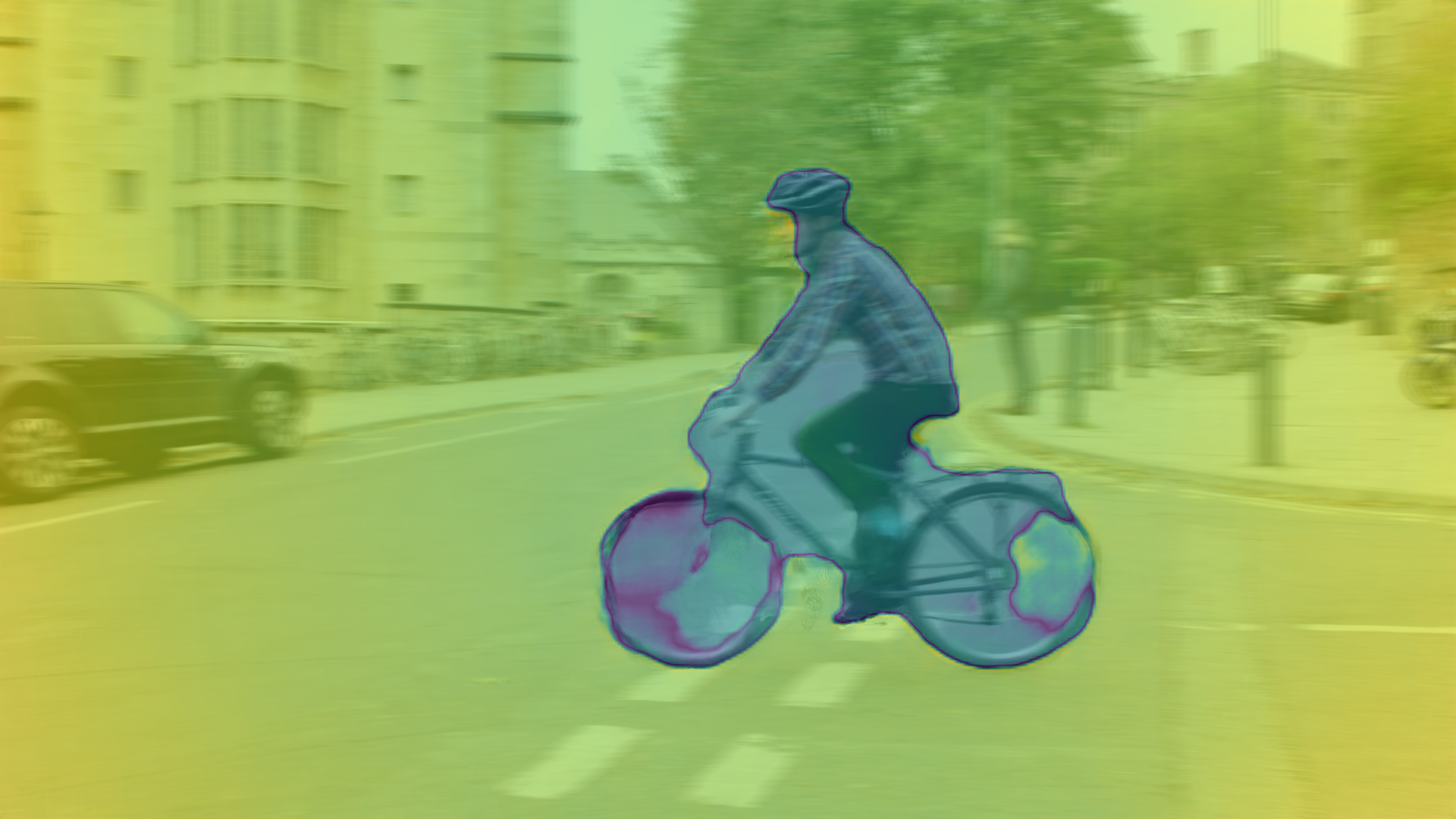}}\;\!\!\!
    \subfloat[Flow diff. weight]{\includegraphics[width=0.330\linewidth]{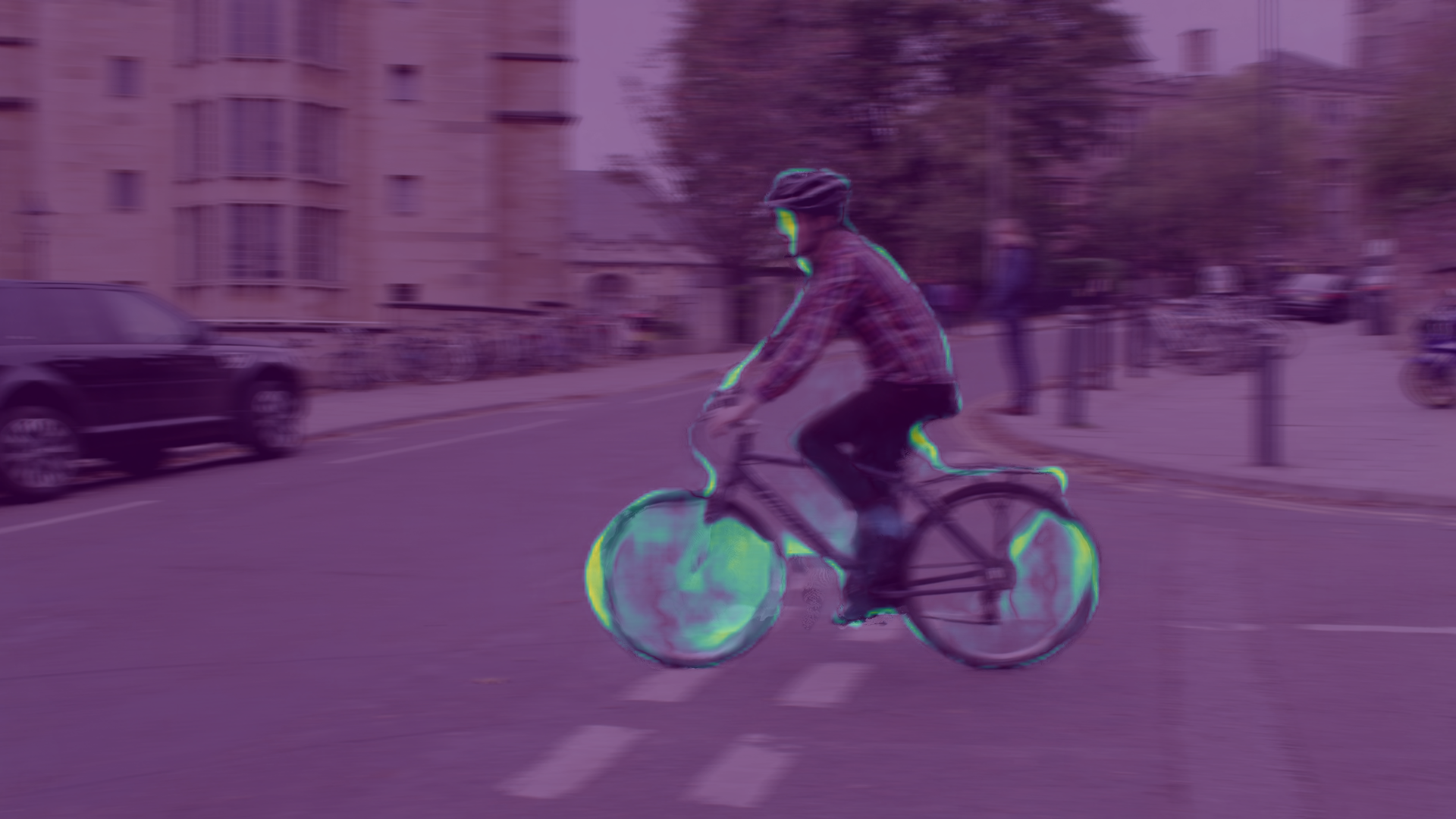}} \\
    \subfloat[Reference]{\includegraphics[width=0.495\linewidth]{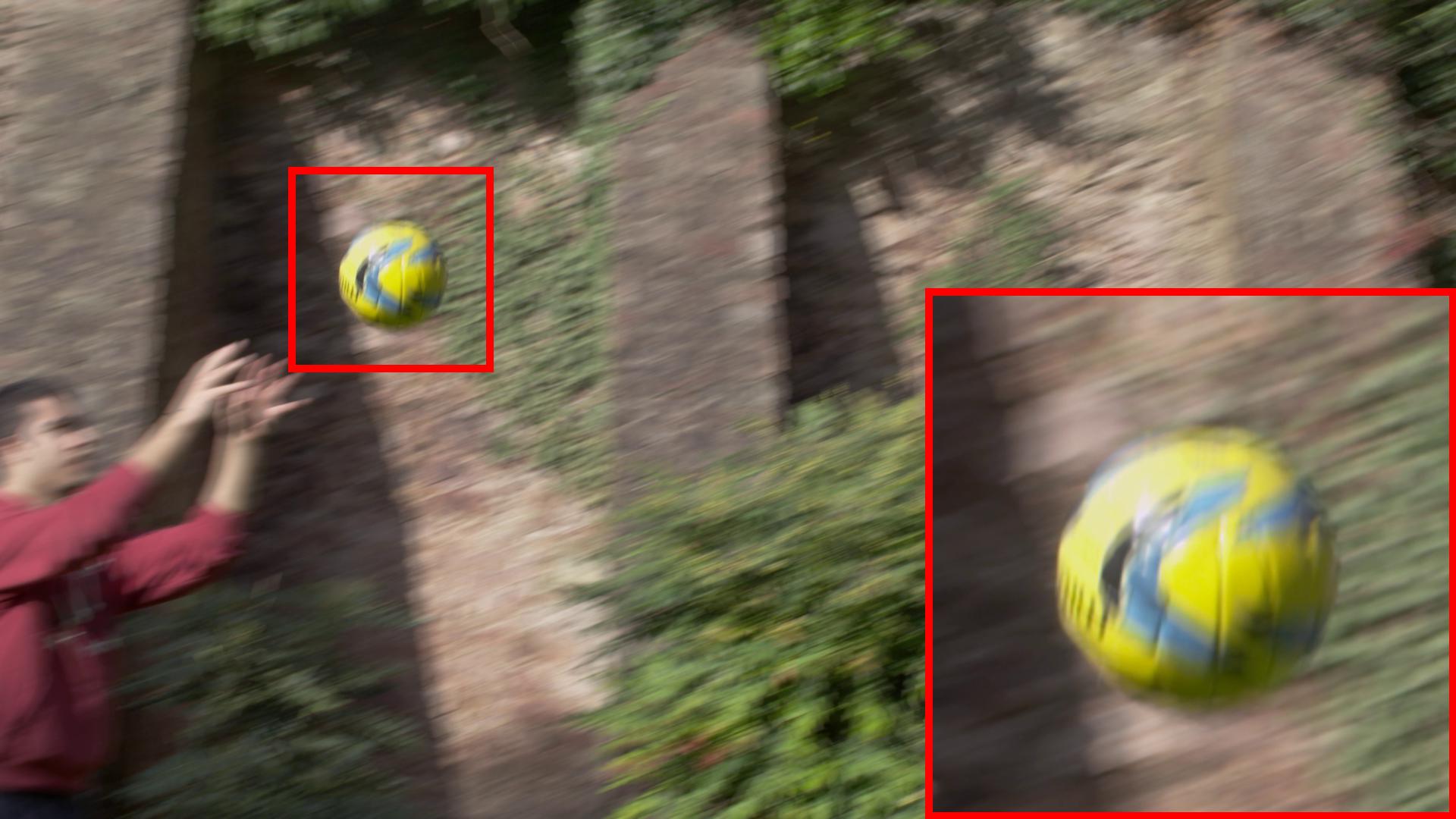}}\;\!\!
    \subfloat[Distorted]{\includegraphics[width=0.495\linewidth]{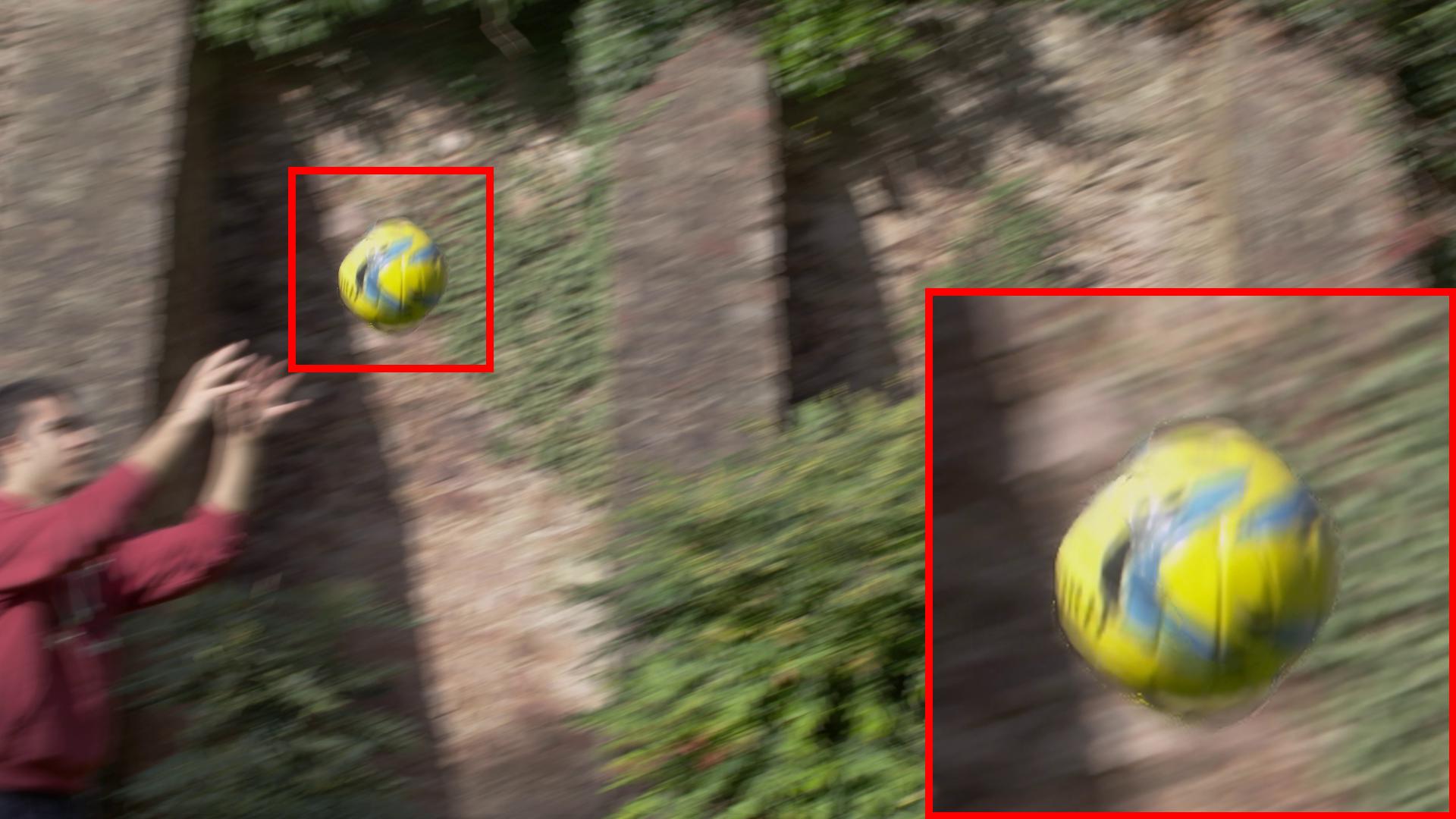}}\\
    \subfloat[Ref. flow weight]{\includegraphics[width=0.330\linewidth]{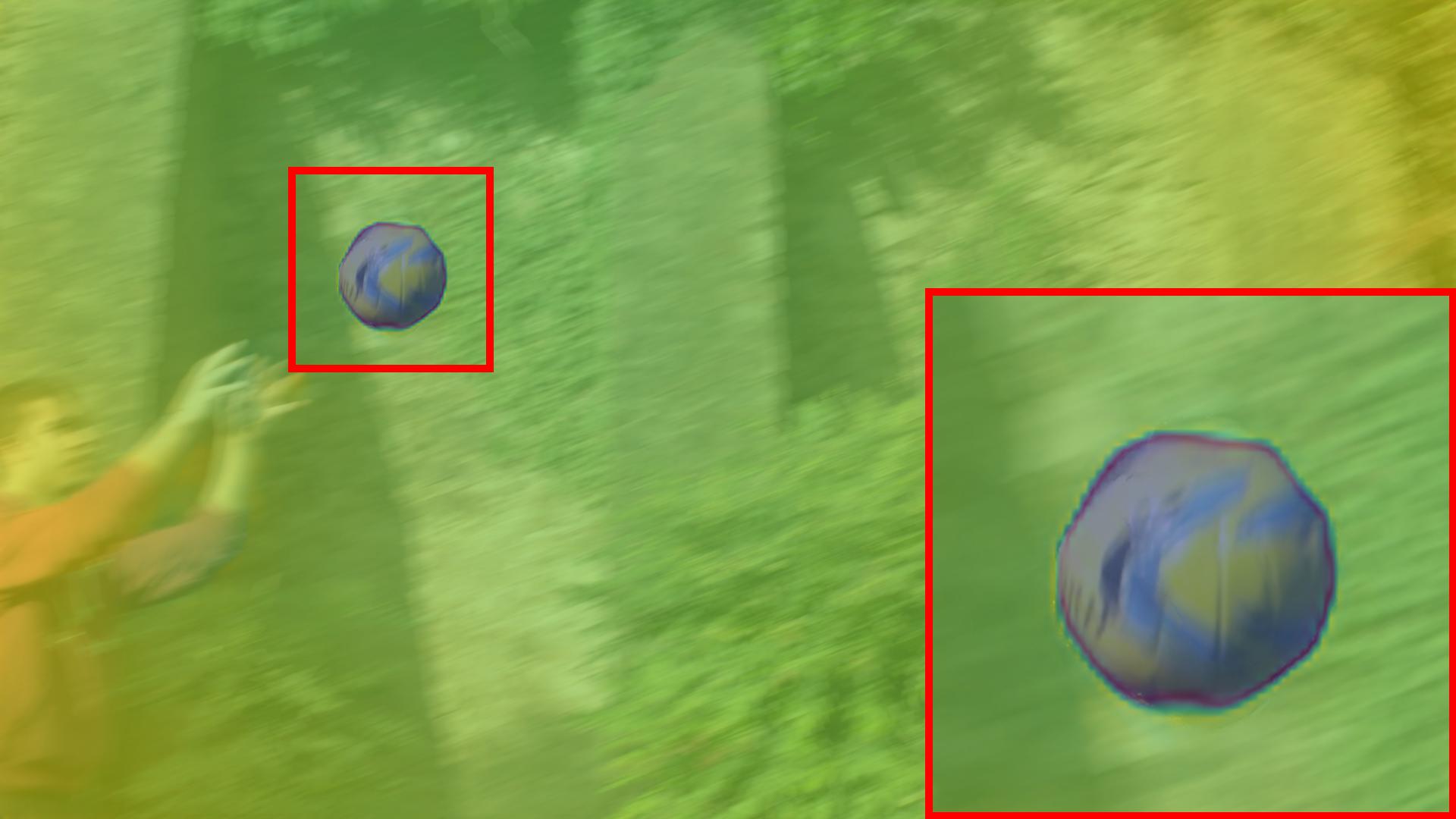}}\;\!\!\!
    \subfloat[Dis. flow weight]{\includegraphics[width=0.330\linewidth]{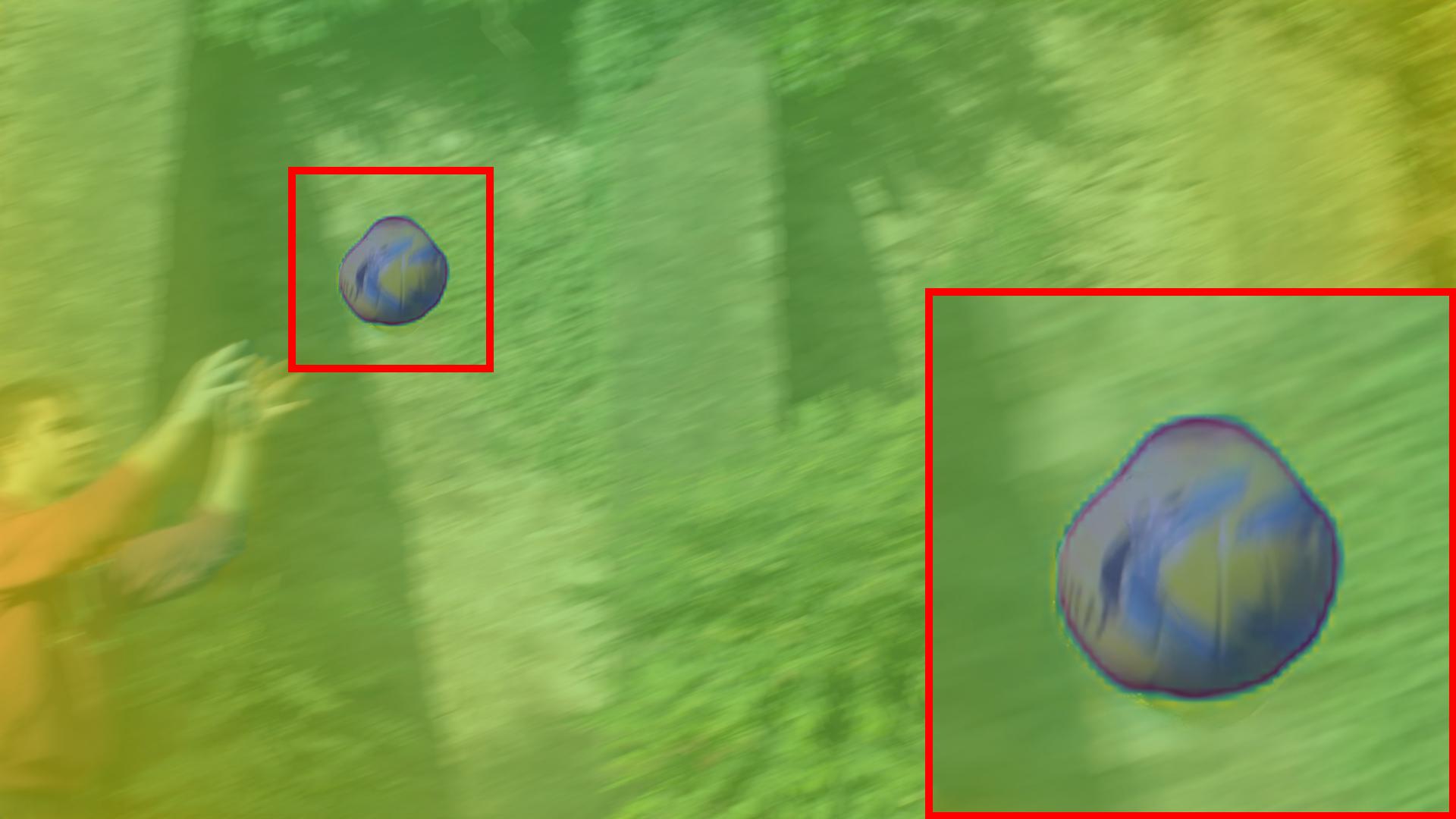}}\;\!\!\!
    \subfloat[Flow diff. weight]{\includegraphics[width=0.330\linewidth]{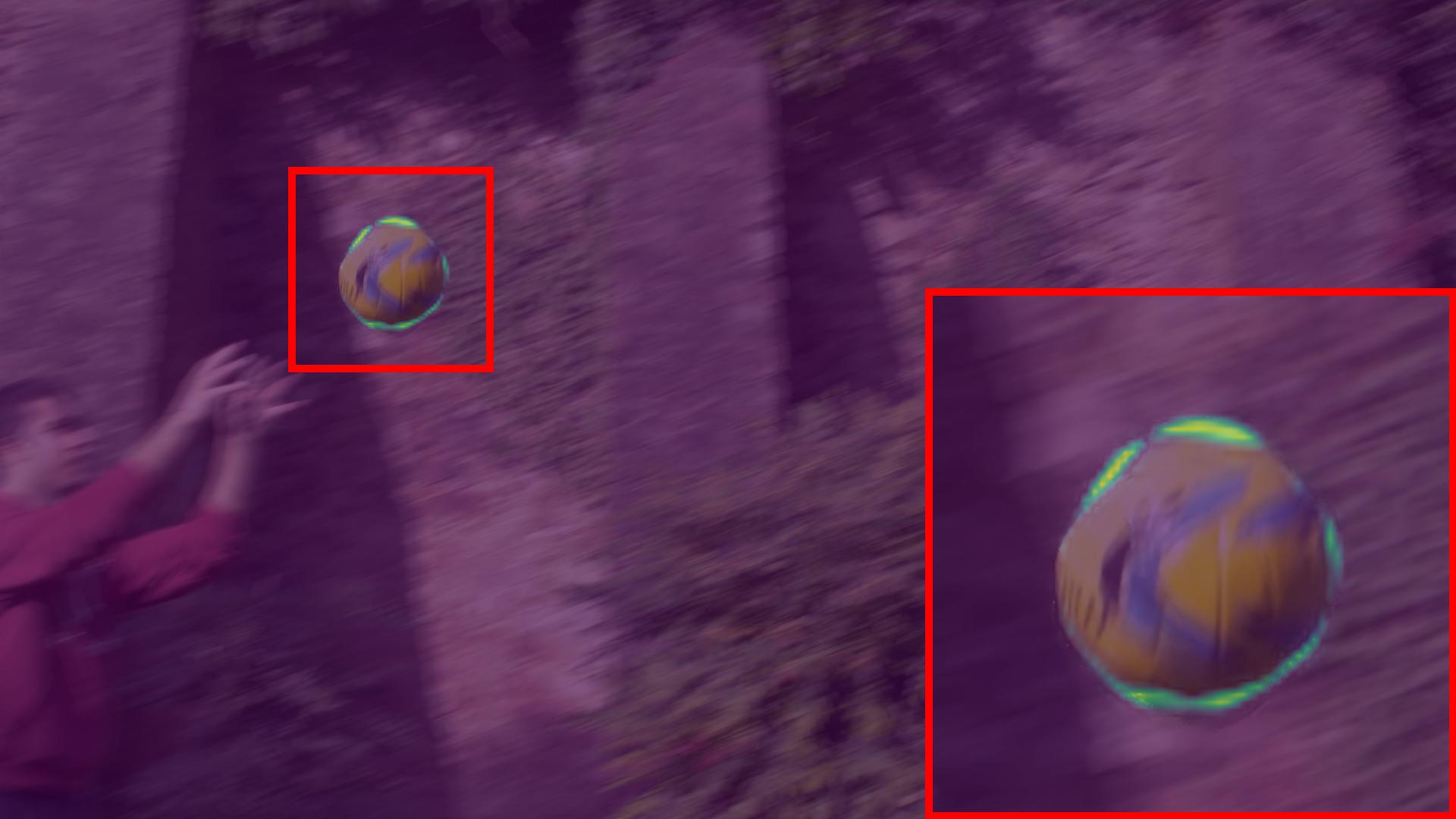}}
    \caption{Visual example demonstrating the advantage of using the proposed flow difference weighting method.}
	\label{fig:ablation}
\end{figure}

\subsection{Model Analysis}
\textbf{Effectiveness of flow difference weighting}. FloLPIPS uses the difference in reference and distorted optical flow maps to perform weighted spatial pooling. To validate this design, we evaluate two more weighting methods by: (i) using the reference flow map and (ii) using the distorted flow map. These are achieved by replacing (\ref{eqn:l2}) with (\ref{eqn:ab1}) and (\ref{eqn:ab2}) respectively.
\begin{gather}
    \Delta F = \norm{F^{ref}}_2 \label{eqn:ab1} \\
    \Delta F = \norm{F^{dis}}_2 \label{eqn:ab2}
\end{gather}
The evaluation results of these variants on BVI-VFI are shown in Table~\ref{tab:ablation}, where it can be noted that the proposed flow difference achieves the best overall performance, further demonstrating the superiority of FloLPIPS. Fig.\ref{fig:ablation} further illustrates the advantage of the proposed weighting method, where the reference and distorted frames are taken from a video with both camera and foreground object motion. It can be observed from Fig.\ref{fig:ablation}(a-b) that distortions are mainly located near the wheels of the bicycle. However, both the weight maps obtained from the reference and distorted flow maps (c-d) focused on the fast-moving background (caused by camera motion), while the flow difference-based weighting (e) successfully captured the interpolation distortions. Similarly in sub-figures (f-j) where the ball undergoing fast motion is being tracked by the camera, the weight map generated using reference and distorted flows failed to capture the salient distortion on the ball, while the proposed method managed to do so.

\textbf{Effect of the flow estimator}. To study the extent to which the proposed method relies on the optical flow estimator employed, we replaced the PWC-Net with two other flow estimators: DISFlow~\cite{kroeger2016fast} and GMFlow~\cite{xu2022gmflow}. The evaluation results of these variants are shown in Table~\ref{tab:flownet}, where it can be observed that the choice of the optical flow estimator does impact the performance of FloLPIPS. For example, when using more recent deep learning (DL)-based PWC-Net and GMFlow, better overall performance was achieved compared to the non-DL method DISFlow. Also, the state-of-the-art GMFlow method results in a slightly higher SROCC value compared to PWC-Net. This implies that more advanced optical flow algorithms can be used in the FloLPIPS framework for VFI quality assessment to achieve better correlation performance with subjective ground truth\footnote{In this work, we use PWC-Net for full evaluation due to the trade off between complexity and performance.}.

\textbf{Differentiablility}. Another advantage of FloLPIPS, besides its superior performance, is that it is fully differentiable, as long as the flow estimator used is so (e.g. PWC-Net). This means that FloLPIPS can potentially (with further complexity reduction) be used as a perceptual loss function for optimising video frame interpolation methods.


\begin{table}[t]
\centering
\caption{Ablation study results on different optical flow estimators.}
\label{tab:flownet}
\begin{tabular}{c|c|c|c}
\toprule
Flow estimator & PLCC $\uparrow$ & SROCC $\uparrow$ & RMSE $\downarrow$ \\
\midrule 
DISFlow & 0.656 & 0.673 & 17.956 \\
GMFlow & 0.680 & 0.695 & 16.091 \\
PWC-Net & 0.706 & 0.683 & 15.546 \\
\bottomrule	
\end{tabular}
\end{table}

\section{Conclusion}\label{sec:conclusion}
In this paper, we introduced FloLPIPS, a full reference video quality assessment method specifically designed for video frame interpolation quality assessment. The proposed method builds upon the popular perceptual image quality metric, LPIPS, and improves its performance by incorporating distortions in the temporal domain represented by discrepancy between the reference and distorted optical flow fields. Such degradation in optical flow maps is used to weight the distortion maps in LPIPS during spatial pooling. FloLPIPS has been quantitatively evaluated and benchmarked against 12 commonly used (or recently reported) quality assessment models on the BVI-VFI database. The results demonstrate that FloLPIPS offers superior performance compared to all tested metrics with statistical significance, while requiring a moderate runtime. The proposed quality metric serves as a better quality assessment tool for VFI applications, and can also be used as a perceptual loss function for training learning-based VFI methods.

\small
\bibliographystyle{IEEEtran}
\bibliography{IEEEexample}

\end{document}